\documentclass[11pt]{article}
\usepackage{fullpage}
\usepackage{amsmath}
\usepackage{amsfonts}
\usepackage[dvips]{epsfig}

\def\##1{\underline{#1}}
\def\=#1{\underline{\underline{#1}}}

\def\+#1{\underline{\bf #1}}
\def\*#1{\underline{\underline{\bf #1}}}

\def\.{\mbox{ \tiny{$^\bullet$} }}

\def\omegao{\omega_{\scriptscriptstyle 0}}

\def\c#1{\cite{#1}}

\renewcommand{\thefootnote}{\fnsymbol{footnote}}

\begin{document}

\begin{center}

\LARGE{ {\bf
Enhanced group velocity in metamaterials}}

\end{center}

\begin{center}

\bigskip
Tom G. Mackay\footnote{Fax: +44 131 650
6553; e--mail: T.Mackay@ed.ac.uk}\\

{\em School of Mathematics,  University of Edinburgh \\
     James Clerk Maxwell Building, The King's Buildings\\
      Edinburgh  EH9 3JZ, UK}
\bigskip

Akhlesh Lakhtakia\footnote{Fax: +1 814 863 7967;
e--mail: axl4@psu.edu}\\

{\em CATMAS --- Computational and Theoretical Materials Sciences Group \\
     Department of Engineering Science and Mechanics \\
     Pennsylvania State University, University Park, PA
     16802--6812, USA}

\end{center}

\renewcommand{\thefootnote}{\arabic{footnote}}
\setcounter{footnote}{0}

\bigskip

\noindent {\bf Abstract.}
 The Bruggeman 
formalism is implemented to estimate the
refractive index of an isotropic, dielectric, homogenized composite
medium (HCM). Invoking
the well--known Hashin--Shtrikman bounds,
we demonstrate
 that the group velocity in  certain HCMs can exceed the group velocities in their
  component materials. Such HCMs should therefore be considered
  as metamaterials.

\bigskip
\noindent
PACS numbers: 41.20.Jb, 42.25.Dd, 83.80.Ab

 \section{Introduction}

By definition, metamaterials exhibit behavior which (i) either their
component materials do not exhibit (ii) or is enhanced relative
to exhibition in the component materials  \cite{Wals}.
Many types of metamaterials may be
conceptualized  through the process
of homogenization \cite{Ward, PSN95,L96},
paving the way for
their realization. For example, a homogenized composite medium (HCM)
may be envisaged which supports the propagation 
of a Voigt wave (which is a planar wave whose amplitude varies linearly
with propagation distance), although
such waves cannot  propagate through its component materials
\cite{ML03a,ML03b}. 

In this communication, we explore the enhancement of
group velocity which may be achieved through homogenization.
 S{\o}lna and Milton
recently considered this issue, by estimating the relative permittivity of
a HCM as the volume--weighted sum of the relative permittivities of
the component materials
 \cite{Solna}. But that estimation is applicable only
 for planar composite materials such as superlattices of thin films,
 and not to the more commonly encountered particulate
 composite materials \cite{Ward, PSN95, L96}. In the following analysis, we implement
the well--established Bruggeman formalism \cite{L96}
 to calculate the effective refractive index of an
isotropic dielectric HCM. Thereby, we  demonstrate  that
metamaterials which support  group velocities exceeding those in
their component materials may be realized as particulate composite materials.

 \section{Analysis}

Consider  a composite
material containing materials labeled $a$ and
$b$, with refractive indexes   $n_a$ and $n_b$, respectively. 
The component materials are envisioned as random distributions of spherical 
particles. Provided that the diameters of these  particles are small compared
with electromagnetic wavelengths, homogenization techniques may be
applied
to estimate the  effective refractive index
of the HCM. 

In particular, the well--established
Bruggeman
homogenization formalism \cite{Ward, L96}~---~which may be 
rigorously derived from 
the strong--permittivity--fluctuation theory \c{TK81, M03}~---~
leads to the equation
\begin{equation}
f_a \frac{n^2_a - n^2_{Br}}{n^2_a + 2 n^2_{Br}} + f_b \frac{n^2_b
- n^2_{Br}}{n^2_b + 2 n^2_{Br}} = 0, \label{Br_eq}
\end{equation}
whose solution yields  $n_{Br}$ as the estimated refractive index of the HCM. 
Here, $f_a$ and $f_b = 1 - f_a$ are the volume
fractions of the component materials. 
In the following,
both component materials
are assumed to have negligible dissipation in the frequency
range of interest.

The group velocity  of a
wavepacket propagating through
the HCM is given as \cite{Jackson}
\begin{equation}
\left.
v_{Br} = \frac{c}{n_{Br}(\omega) + \omega \frac{d n_{Br}}{d
\omega}} \; \right|_{\omega(k_{avg})}, \hspace{25mm}
\end{equation}
where $v_{Br}$ is evaluated at the angular frequency $\omega = 
\omega (k_{avg})$,  with $k_{avg}$ being the average wavenumber of the wavepacket,
and $c$ is the speed of light in free space.
Similarly, the respective group velocities in component materials $a$ and
$b$ are given by
\begin{equation}
\left.
v_{\ell} = \frac{c}{n_{\ell}(\omega) + \omega \frac{d n_{\ell}}{d
\omega}}  \; \right|_{\omega(k_{avg})}, \hspace{25mm} \left( \ell = a,b \right).
\end{equation}
We proceed to establish upper and lower bounds on  $v_{Br}$, in
terms of $n_a$ and $n_b$. In
particular, we demonstrate that the inequalities
\begin{equation}
v_{Br} > v_{\ell}\,, \hspace{25mm} \left( \ell = a,b \right)
\label{cond1}
\end{equation}
can be satisfied for certain values of $n_a\geq 1$, $ n_b \geq 1$,
$\frac{d n_{a}}{d \omega}>0$, and $\frac{d n_{b}}{d \omega} > 0$.

Differentiation of  both sides of Eq. (\ref{Br_eq}) with respect to $\omega$ yields
\begin{equation}
\frac{d n_{Br}}{d \omega} = \delta_a \frac{d n_a}{d \omega} +
\delta_b \frac{d n_b}{d \omega}, \label{dndw}
\end{equation}
where
\begin{equation}
\left.
\begin{array}{l}
 \delta_a =  \displaystyle \frac{f_a n_a
n_{Br} \left( n^2_b + 2 n^2_{Br} \right)^2}{ f_a n^2_a \left(
n^2_b + 2
n^2_{Br} \right)^2 + f_b n^2_b \left( n^2_a + 2 n^2_{Br} \right)^2} \\
\vspace{-4mm} \\  \delta_b  = \displaystyle \frac{f_b n_b n_{Br}
\left( n^2_a + 2 n^2_{Br} \right)^2}{ f_a n^2_a \left( n^2_b + 2
n^2_{Br} \right)^2 + f_b n^2_b \left( n^2_a + 2 n^2_{Br}
\right)^2} \label{delta}
\end{array}
\right\}.
\end{equation}
Upper and lower  bounds on $\delta_a$ and $\delta_b$ may be
established by exploiting the Hashin--Shtrikman bounds $n_L$
and $n_U$  on
$n_{Br}$ \cite{H-S}; i.e.,
\begin{equation}
n_L < n_{Br} < n_U, \label{HS}
\end{equation}
where
\begin{equation}
\left.
\begin{array}{l}
n^2_L = n^2_b + \displaystyle \frac{3 f_a n^2_b \left( n^2_a -
n^2_b \right)}
{n^2_a + 2 n^2_b - f_a \left( n^2_a - n^2_b \right)}  \\
\vspace{-4mm} \\ n^2_U = n^2_a + \displaystyle \frac{3 f_b
n^2_a \left( n^2_b - n^2_a \right)} {n^2_b + 2 n^2_a - f_b \left(
n^2_b - n^2_a \right)}
\end{array}
\right\} . \label{HS+}
\end{equation}
Combining Eqs. (\ref{delta})--(\ref{HS+}), we get
\begin{equation}
 \rho_\ell < \delta_\ell <
\kappa_\ell\,, \hspace{25mm} \left( \ell = a,b \right),
\end{equation}
where
\begin{equation}
\left.
\begin{array}{l}
\kappa_a = \displaystyle \frac{f_a n_a n_U \left( n^2_b + 2
n^2_U \right)^2}{ f_a n^2_a \left( n^2_b + 2 n^2_L \right)^2 + f_b
n^2_b \left( n^2_a + 2 n^2_L \right)^2}
\\ \vspace{-4mm} \\ \kappa_b =  \displaystyle \frac{f_b n_b n_U \left( n^2_a + 2 n^2_U \right)^2}{ f_a
n^2_a \left( n^2_b + 2 n^2_L \right)^2 + f_b n^2_b \left( n^2_a +
2 n^2_L \right)^2}
\end{array}
\right\}
 \label{tau}
\end{equation}
and
\begin{equation}
\left.
\begin{array}{l}
\rho_a =  \displaystyle \frac{f_a n_a n_L \left( n^2_b + 2
n^2_L \right)^2}{ f_a n^2_a \left( n^2_b + 2 n^2_U \right)^2 + f_b
n^2_b \left( n^2_a + 2 n^2_U \right)^2}
\\ \vspace{-4mm} \\ \rho_b = \displaystyle \frac{f_b n_b n_L \left( n^2_a + 2 n^2_L \right)^2}{ f_a
n^2_a \left( n^2_b + 2 n^2_U \right)^2 + f_b n^2_b \left( n^2_a +
2 n^2_U \right)^2}
\end{array}
\right\}.
 \label{beta}
\end{equation}
Thus, we have
\begin{equation}
\rho_a \frac{d n_a}{d \omega} + \rho_b \frac{d n_b}{d \omega} <
\frac{d n_{Br}}{d \omega} < \kappa_a \frac{d n_a}{d \omega} +
\kappa_b \frac{d n_b}{d \omega},
\end{equation}
and the  group velocity in the HCM is
accordingly  bounded as
\begin{equation}
 v_L < v_{Br}  <  v_U\,,
\end{equation}
with
\begin{equation}
\left.
\begin{array}{l}
v_L = \displaystyle \frac{c}{n_{U} + \omega \left( \kappa_a
\frac{d n_a}{d \omega} + \kappa_b \frac{d n_b}{d \omega} \right)}
\\ \vspace{-4mm} \\ v_U = \displaystyle \frac{c}{n_{L} + \omega \left( \rho_a
\frac{d n_a}{d \omega} + \rho_b \frac{d n_b}{d \omega} \right)}
\end{array}
\right\}.
\end{equation}

If  the inequalities
\begin{equation}
\left.
\begin{array}{l}
\displaystyle n_{U} + \omega \left( \kappa_a \frac{d n_a}{d
\omega} + \kappa_b \frac{d n_b}{d \omega} \right) < n_a +
\omega \frac{d n_a}{d \omega} \\
\vspace{-4mm} \\
\displaystyle n_{U} + \omega \left( \kappa_a \frac{d n_a}{d
\omega} + \kappa_b \frac{d n_b}{d \omega} \right) < n_b + \omega
\frac{d n_b}{d \omega} \label{cond2}
\end{array}
\right\}
\end{equation}
hold for certain component materials, then
 the inequalities (\ref{cond1}) are
 automatically  satisfied. 
 
 The inequalities (\ref{cond2})  reduce to the particularly simple inequality 
\begin{equation}
 n_{U} + \omega  \frac{d n_a}{d
\omega} \left( \kappa_a + \kappa_b - 1 \right) + \kappa_b \left(
n_a - n_b \right) < n_a,   \label{cond_fin}
\end{equation}
if $v_a = v_b$. The
conditions
\begin{equation}
\left.
\begin{array}{lcl}
 n_{U}  + \kappa_b \left( n_a - n_b \right) &<&
n_a \\
\vspace{-4mm} \\ \kappa_a + \kappa_b - 1& > & 0
\end{array}
\right\}
\end{equation}
are  satisfied, for example, by $n_a =3$, $n_b = 1.2$ and $f_a =
0.9$. Thus, the inequality  (\ref{cond_fin}) holds, provided that the
dispersive term  $\frac{d n_a}{d \omega}$ is sufficiently small.

 \section{Numerical results}

Let us illustrate the phenomenon represented by the inequalities
(\ref{cond1})
 by means of
 a
specific numerical example. Consider a particulate composite material
at a particular value $\omegao$
of  $\omega$. At the chosen angular frequency, let
 $n_a = 5$, $n_b = 1.2$,
 $ \frac{d n_a}{d \omega}\Big\vert_{\omega=\omegao} =
0.5 / \omegao $, and $\frac{d n_b}{d \omega}\Big\vert_{\omega=\omegao} = 5.5 / \omegao$.
Significantly,  material $a$ has a high refractive index
but low dispersion in the neighbourhood of $\omegao$, whereas   high
dispersion in material $b$ is combined with  a low refractive index.

 The Bruggeman estimate of the refractive index of
the HCM, namely $n_{Br}$, is
  plotted as a function of the volume fraction $f_a$ in figure~1.
Also shown are the upper and lower Hashin--Shtrikman bounds,  $n_U$ and $n_L$, on
$n_{Br}$,  as well as the
parameters $\delta_a$ and $\delta_b$. The Bruggeman estimate
adheres closely to the lower bound $n_L$ at low values of
$f_a$, whereas at high values of $f_a$ the difference
between $n_{Br}$ and its upper bound $n_U$ becomes marginal.
The observed agreement between $n_{Br}$ and $n_L$ at  low
$f_a$ reflects the fact that the lower
Hashin--Shtrikman bound  is equivalent to the Maxwell
Garnett estimate of the refractive index of the HCM
 arising from spherical particles of material $a$
embedded in the host material $b$ \cite{L96}. The Maxwell Garnett estimate is
only valid then at low values of $f_a$. 
As the volume fraction becomes increasingly small, the 
 Bruggeman estimate ($n_{Br}$) and the 
Maxwell Garnett estimate (low $f_a$ value of $n_L$) converge on $n_b$.
In a similar manner, the
agreement between  $n_{Br}$ and $n_U$ at high values of $f_a$
is indicative of the fact that the upper
  Hashin--Shtrikman bound   is equivalent to the Maxwell
Garnett estimate of the refractive index of the HCM
 arising from spherical particles made of material $b$  embedded in 
host material $a$;  the Maxwell Garnett estimate then holds only at
high values of $f_a$. In the limit $f_a \rightarrow 0$,
the coefficients $\delta_a \rightarrow 0$ and $\delta_b
\rightarrow 1$;  while $\delta_a \rightarrow 1$ and $\delta_b
\rightarrow 0$ as $f_a \rightarrow 1$.

   The corresponding 
group velocities  $v_a$, $v_b$
   and $v_{Br}$ are plotted as functions of   $f_a$ in figure~2. The upper and lower bounds on $v_{Br}$
as given by $v_U$ and $v_L$, respectively, are also
displayed.
 Clearly,  we have $v_{Br} > v_a$ and
$v_{Br} > v_b$ for $f_a > 0.67$.

The inequalities  (\ref{cond1}) hold only over  a relatively small
range of parameter values. For example, the phase space in which
the inequalities (\ref{cond1}) are satisfied is illustrated in figure~3 for $n_a = 5$,
$ \frac{d n_a}{d \omega}\Big\vert_{\omega=\omegao} = 0.5 / \omegao$ and  $f_a = 0.8$. With
these  relationships fixed for the component material $a$,
 we find that $v_{Br} > v_a$ and $v_{Br} > v_b$ 
for
\begin{itemize}
\item[(i)] $1.17 < n_b <
1.23$ with $ \frac{d n_b}{d \omega}\Big\vert_{\omega=\omegao} = 5.51 / \omegao$;  and 
\item[(ii)]
$ 5.45 / \omegao < \frac{d n_b}{d \omega}\Big\vert_{\omega=\omegao} < 5.57 / \omegao
$ with $n_b = 1.2$.
\end{itemize}

\section{Concluding remarks}

We conclude that  the group velocity in an isotropic,
dielectric, particulate composite material~---~as estimated via the Bruggeman homogenization
formalism~---~can exceed the  group velocities in its component
materials. This metamaterial
characteristic may be achieved through homogenizing (i) a
component material $a$ with high refractive index and
low dispersion with (ii) a component material $b$ with low refractive
index and high dispersion. Neither anomalous dispersion nor an
explicit frequency--dependent model of the refractive index (unlike Ref. \cite{Solna})  is
required to demonstrate this characteristic.

Improved estimates of HCM group velocity may be achieved
through the implementation of homogenization approaches which 
take into better  account  the distributional statististics
of the component materials (e.g., the
strong--permittivity--fluctuation theory 
approach \c{TK81, M03}). In particular, the effects of
coherent scattering losses~---~which are neglected in the 
present study~---~may well result in a moderation of the 
 group velocity. 
Such studies are currently being undertaken, especially
in light of the recent emergence of 
metamaterials wherein the phase velocity and the time--averaged Poynting
vector are oppositely directed \c{LMW03}.

\vspace{10mm}

\noindent {\bf Acknowledgements.} TGM acknowledges the
financial support of \emph{The Nuffield Foundation}. AL thanks
the Trustees of the Pennsylvania State University for a
sabbatical leave of absence.

\newpage

\section*{List of Figure Captions}

\noindent Fig. 1. The estimated refractive index $n_{Br}$ (solid line), the
 upper and lower Hashin--Shtrikman bounds on $n_{Br}$ (broken
 dashed lines, labeled as $n_U$ and $n_L$),
and the coefficients $\delta_a$ and $\delta_b$
 (dashed lines), all plotted as functions of the volume fraction $f_a$, when
 $n_a=5$ and $n_b=1.2$.

\noindent Fig. 2. The estimated group velocity $v_{Br}$ (solid line) and its
upper and
  lower bounds (broken dashed lines, labeled as $v_U$ and $v_L$),
 along with the   group velocities
   $v_a$ and $v_b$ (broken dashed lines) in the component materials,
plotted as functions of the volume fraction $f_a$, when
 $n_a=5$, $n_b=1.2$, $\frac{dn_a}{d\omega}\Big\vert_{\omega=\omegao} = 0.5/\omegao$ and
 $\frac{dn_b}{d\omega}\Big\vert_{\omega=\omegao}=5.5/\omegao$. All group velocities
are normalized with respect to $c$.

\noindent Fig. 3.   The shaded region indicates the portion of the $\alpha$--$\beta$ phase
space wherein $v_{Br}
> v_a$ and $v_{Br} > v_b$; here,
$\alpha = \frac{n_a}{n_b}$
 and $\beta = \Big(\frac{d
n_a}{d \omega} / \frac{d n_b}{d \omega}\Big)\Big\vert_{\omega=\omegao}$. This region was demarcated for
$n_a = 5$, $ \frac{d n_a}{d \omega}\Big\vert_{\omega=\omegao} = 0.5 / \omegao$ and
 $f_a = 0.8$.

\newpage

  \begin{figure}[h]\centerline{\scalebox{1}{\includegraphics{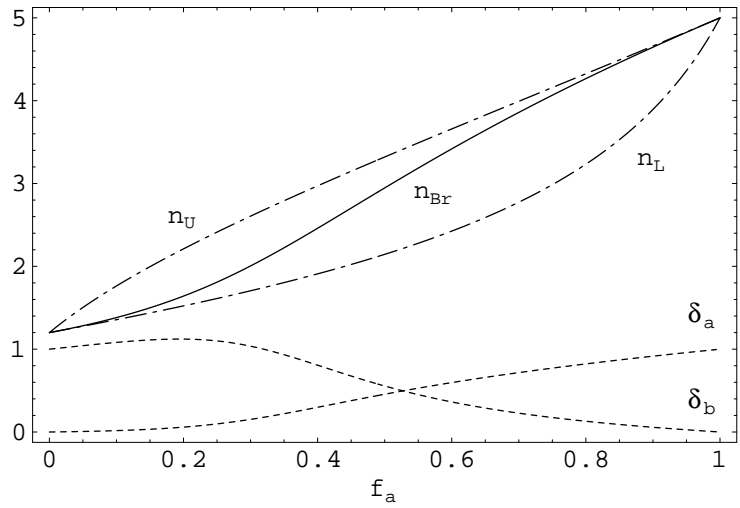}}}
  \caption{The estimated refractive index $n_{Br}$ (solid line), the
 upper and lower Hashin--Shtrikman bounds on $n_{Br}$ (broken
 dashed lines, labeled as $n_U$ and $n_L$),
and the coefficients $\delta_a$ and $\delta_b$
 (dashed lines), all plotted as functions of the volume fraction $f_a$ when
 $n_a=5$ and $n_b=1.2$.
  }
   \label{fig1}
  \end{figure}

\newpage

 \begin{figure}[h]\centerline{\scalebox{1}{\includegraphics{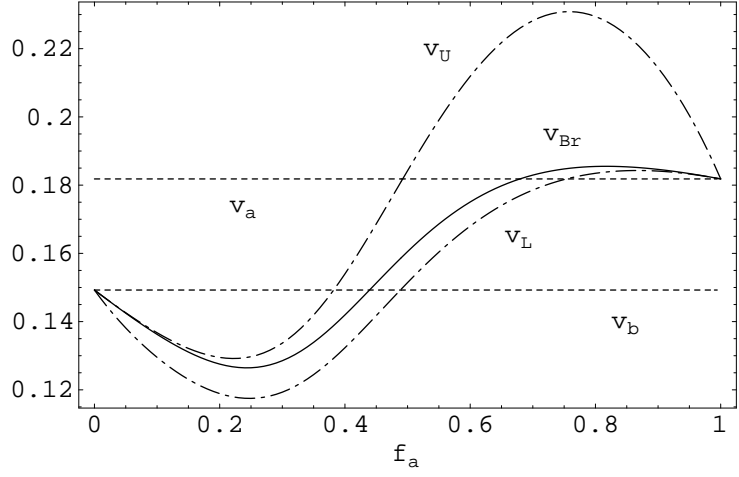}}}
  \caption{The estimated group velocity $v_{Br}$ (solid line) and its
upper and
  lower bounds (broken dashed lines, labeled as $v_U$ and $v_L$),
 along with the   group velocities
   $v_a$ and $v_b$ (broken dashed lines) in the component materials,
plotted as functions of the volume fraction $f_a$, when
 $n_a=5$, $n_b=1.2$, $\frac{dn_a}{d\omega}\Big\vert_{\omega=\omegao} = 0.5/\omegao$ and
 $\frac{dn_b}{d\omega}\Big\vert_{\omega=\omegao}=5.5/\omegao$. All group velocities
are normalized with respect to $c$.   }
  \label{fig2}
  \end{figure}

\newpage

 \begin{figure}[h]\centerline{\scalebox{1}{\includegraphics{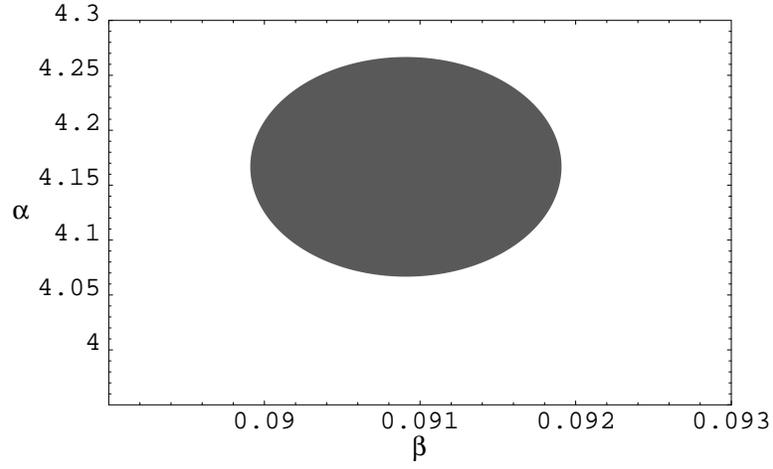}}}
  \caption{The shaded region indicates the portion of the $\alpha$--$\beta$ phase
space wherein $v_{Br}
> v_a$ and $v_{Br} > v_b$; here,
$\alpha = \frac{n_a}{n_b}$
 and $\beta = \Big(\frac{d
n_a}{d \omega} / \frac{d n_b}{d \omega}\Big)\Big\vert_{\omega=\omegao}$. This region was demarcated for
$n_a = 5$, $ \frac{d n_a}{d \omega}\Big\vert_{\omega=\omegao} = 0.5 / \omegao$ and
 $f_a = 0.8$.}
  \label{fig3}
  \end{figure}

\end{document}